\documentstyle{article}
\title {
IMPROVEMENT OF EFFICIENCY IN GENERATING RANDOM $U(1)$ VARIABLES
WITH BOLTZMANN DISTRIBUTION
       }
\author{
Tetsuya Hattori \\
{ \small Faculty of Engineering, Utsunomiya University,
Ishii-cho, Utsunomiya, Tochigi 321, Japan
} \\
{ \small  e-mail address: hattori@tansei.cc.u-tokyo.ac.jp} \\
\and
Hideo Nakajima \\
{ \small Faculty of Engineering, Utsunomiya University,
Ishii-cho, Utsunomiya, Tochigi 321, Japan } \\
{ \small  e-mail address: nakajima@kinu.infor.utsunomiya-u.ac.jp} \\
       }
\date{\today}

\begin{document}
\maketitle
\begin{abstract}
A method for generating random $U(1)$ variables with Boltzmann
distribution is presented.
It is based on the rejection method with transformation of variables.
High efficiency is achieved for
all range of temparatures or coupling parameters,
which makes the present method especially suitable for
parallel and pipeline vector processing machines.
Results of computer runs are presented to illustrate the efficiency.
An idea to find such algorithms is also presented,
which may be applicable to other distributions of interest in
Monte Carlo simulations.
\bigskip

\newline

\end{abstract}

\section{Introduction.}  \label{s1}
%section 1
Monte Carlo numerical integration method, or Monte Carlo simulation,
has been widely used
in the numerical study of quantum field theroies with lattice formalism and
statistical mechanics of spin systems.
In performing Monte Carlo simulations,
one must generate sequences of random numbers
with given probability distributions.
Each random number is used to ^^ update' a spin or a gauge variable.
Probability distributions which appear in such calculations
have parameter dependences.
These parameters carry the information of temperature and other
thermodynamic quantities,
and also that of the neighboring spin states.

For a fixed probability distribution
there are algorithms (see, for example, \cite{T} and references therein)
which may be very efficient in generating random numbers.
In Monte Carlo simulations, however,
we have fluctuations in the neighboring spins,
which results in changes of the parameters within a single program.
One faces the problem of finding an algorithm
suitable to a class of distributions parametrized by the parameters.
These parameters are changed over a wide range,
so that we must find a method which maintain high efficiency
for all range of parameters,
especially in the limit where a parameter tends to $\infty$\
and the distribution becomes singular.

Another aspect which we consider, is the efficiency in parallel or
pipeline vector processing.
As far as we know, much of the currently available vector processors
work efficiently when there are no ^^ if-branches' in the
program.

We study a method based on a
rejection method combined with a change of variables \cite{B},
which is an approach that is widely used.
In the case of rejection method,
to avoid ^^ if-branches' we have to fix the number of
iterations of the rejection trials.
It is particularly important to have high acceptance rate
{\it uniformly\/} in the parameters.

The aim of the present study is to find a suitable method
for generating a sequence of random $U(1)$ numbers
which we use to update site or link variables of a canonical ensemble
for $U(1)$ lattice gauge theories or $U(1)$ spin sytems,
{}from the point of view mentioned above.
An example of a program for our proposal is given in appendix~\ref{sd}.

In section~\ref{s2} we set up the problem.
In section~\ref{s3} we give our strategy for the solution,
and in section~\ref{s4} we give an answer to the problem.
We present the results of our efficiency tests of the algorithm
in section~\ref{s5}.

\section{Random $U(1)$ variable and rejection method.}  \label{s2}
%section 2
In the following we call a sequence of random $U(1)$ numbers,
a random $U(1)$ variable:
A random $U(1)$ variable is a sequence of numbers
(the angle variables)
\begin{equation}
\label{e21}
\theta_{1}\,,\ \theta_{2}\,,\ \theta_{3}\,,\cdots,
\end{equation}
whose distribution
\[ P([\theta,\theta+d\theta])=f_{a}(\theta)\,d\theta \]
is
given by the density function
\begin{eqnarray*}
f_{a}(\theta) &=& N_{a} \exp(a \cos(\theta-\theta_{0}))\,,\
\end{eqnarray*}
where $N_{a}$ is a normalization constant.
In the practical applications, the parameter $a$ is
proportional to the inverse temperature $1/T$
or the inverse coupling $1/g^{2}$,
and both $a$ and the constant $\theta_{0}$\
contain the effect of interactions with other sites or link variables.
By the shift of variable
$\theta'= \theta-\theta_{0}\ \mbox{if}\ a>0\,,$ and
$\theta'= \theta-\theta_{0}-\pi\ \mbox{if}\ a<0\,,$
we may assume without loss of generality that $\theta_{0}=0$
and $a \ge 0\,.$
Hence
\begin{eqnarray}
f_{a}(\theta) &=& N_{a} \exp(a \cos \theta)\,,\ \
-\pi\leq \theta<\pi\,,\ a> 0\,, \nonumber
\\
\label{e27}
\frac{1}{N_{a}} &=& \int_{-\pi}^{\pi} \exp(a \cos\theta)\,d\theta\,.
\end{eqnarray}
The right hand side of eq.~(\ref{e27}) is (modulo constant)
an integral representation of modified Bessel function $I_0(a)\,.$

On computers we start with uniform random variables
\begin{equation}
\label{e22}
\omega_{1}\,,\ \omega_{2}\,,\ \omega_{3}\,,\cdots,
\end{equation}
with the probability distribution
\[ P([\omega,\omega+d\omega])=d\omega\,,\ 0\leq\omega<1\,. \]
If we know an expression for a function $X(\omega)$
expressible as a computer program of small time consumption
such that the sequence
\[
\theta_{i}=X(\omega_{i})\,,\ i=1,2,3,\cdots,
\]
is the random $U(1)$ variable (\ref{e21}),
then there is in principle no problem.
Such a function $X(x)$ is formally given by
\begin{eqnarray*}
X(x)&=&F^{-1}(x)\,;
\\
F(t)&=&P([-\pi\leq\theta<t])
=\int_{-\pi}^{t}f_{a}(\theta)\,d\theta\,.
\end{eqnarray*}
(Here and in the following, $F^{-1}$ is the inverse function of $F\,;$
$F^{-1}(F(x))=x\,.$)
Unfortunately,
we do not know a suitable expression of $X(x)$ for random $U(1)$ variables.
(Note that we have the parameter $a$ dependence.)

Usually the rejection method is adopted to solve the problem.
The rejection method, combined with transformation of variables,
is defined as follows.
Let $\tilde{f}(\theta)$ be some approximate density function to the
density function $f_{a}(\theta)\,.$
We assume that the density functions are continuous.
Note that they are normalized to satisfy
${\displaystyle \int_{-\pi}^{\pi} f(\theta)\,d\theta=1 \,.}$
Suppose that there is a monotonic function $h$ which satisfies
\[ h(0)=-\pi\,,\ \ h(1)=\pi\,, \]
and
\begin{equation}
\label{e23}
\tilde{f}(h(x))\,h'(x)=1\,,\ 0<x<1\,,
\end{equation}
where $h'$ is the derivative of $h\,,\
{\displaystyle h'=\frac{dh}{dx}}\,.$
(For the moment, we suppress possible parameter dependences
of $\tilde{f}$\ and $h\,.$)
Define a function $g$ by
\begin{eqnarray}
\label{e24}
g(x) &=& R(a)\,\frac{f_{a}(h(x))}{\tilde{f}(h(x))}\,,\ 0\leq x<1\,,
\\
\label{e25}
R(a) &=& \min_{-\pi\leq \theta<\pi}
\left\{ \frac{\tilde{f}(\theta)}{f_{a}(\theta)} \right\} \,.
\end{eqnarray}
Let $\omega_{j}$ and $\omega'_{j}$ with $j=1,2,3,\cdots,$
be two sequences of independent uniform random variables as in (\ref{e22}).
Define a subsequence
\[ \tilde{\omega}_{i}=\omega_{j_{i}}\,,\ i=1,2,3,\cdots\,, \]
of the sequence $\{\omega_{j}\}$
by selecting the numbers $j=j_{i}$ that satisfy
\[ \omega'_{j}\leq g(\omega_{j})\,. \]
Then the sequence
\begin{equation}
\label{e26}
h(\tilde{\omega}_{1})\,,\ h(\tilde{\omega}_{2})\,,\ h(\tilde{\omega}_{3})\,,\
\cdots,
\end{equation}
is the random $U(1)$ variable we are looking for.
We call the rate of picking up $\tilde{\omega_{i}}$ out of $\omega_{j}$
the acceptance rate.
To achieve high efficiency, the acceptance rate should be high.
It can be shown that the acceptance rate is equal to $R(a)$
in eq.~(\ref{e25}).
The proof that the distribution of (\ref{e26}) is the one
with the density function $f_{a}\,,$
and the proof that the acceptance rate is equal to $R(a)$,
are given in Appendix~\ref{sa}.

Note that the density functions are non-negative functions and that
the integrated values of $\tilde{f}$\ and $f_{a}$ are normalized to unity,
hence $0\leq R(a)\leq1\,,$ which should hold for an acceptance rate.
Note also that
if $\tilde{f}$\ is a good approximation to $f_{a}\,,$
then the function $g$ is almost flat.

Our problem is to find a good $\tilde{f}\,.$

\section{Approximate distributions and the optimization of
the acceptance rate.}  \label{s3}
%section 3
In order to keep high efficiency,
we must choose $\tilde{f}$ with high acceptance rate $R(a).$
To illustrate the implications of this statement,
let us first consider the simplest choice of $\tilde{f}\,,$
the flat distribution.
Hereafter, we refer to this choice as the ^^ direct' method.

The ^^ direct' method is defined by choosing $\tilde{f}$ in
eq.~(\ref{e23}) to be a constant function;
\begin{eqnarray}
\tilde{f}(\theta) &=& \frac{1}{2 \pi}\,, \\
h(x) &=& (2 x -1) \pi\,.
\end{eqnarray}
The acceptance rate is
\begin{eqnarray}
R(a)=\frac{1}{2 \pi N_{a} \exp(a)}\,.
\end{eqnarray}
For $a$ near zero (^^ high temperature'), the acceptance rate is high;
$\displaystyle R(a) \approx 1-a+\frac{3}{4} a^2\,,\ a \ll 1\,,$
while for large $a$ (^^ low temperature'), the acceptance rate
becomes very small;
\[
R(a) \approx \frac{1}{\sqrt{2 \pi a}} \,,\ a \gg 1\,.
\]
In the limit $a \to \infty\,,$ the acceptance rate approaches zero.
We have very slow effective generation of
random $U(1)$ variables
for large values of $a$ with the ^^ direct' method.
In other words, the improvement of the efficiency of
generating random $U(1)$ variables, which we measure by time lapse
per a random variable generation,
depends basically on the improvement of the acceptance rate.
The reason that the acceptance rate is small for large $a$ is
that for large $a$ the original distribution $f_{a}$ has a large peak
at $\theta=0\,.$ The distribution becomes highly non-uniform and
the flat (uniform) distribution $\tilde{f}$ is not a good approximation.

As noted in the section~\ref{s1} (see also section~\ref{s5}),
it is even more important
to keep high acceptance rate uniformly in $a$
when we use parallel processing machines.

Our aim is to find a $\tilde{f}$ which is a good approximation
to the original distribution $f_{a}$
for all values of $a\,,$
in particular, the $a \to \infty$ limit.
In a sense, this is to find a family of distributions
which interpolates the flat distribution ($a=0$) and
the delta function distribution ($a=\infty$)
expressible as a simple computer program.

For large $a$, the density function $f_{a}$ has a sharp
peak at $\theta=0\,.$ Therefore $\tilde{f}$\
should be a very good approximation at $\theta=0\,,$
while being not too bad an approximation at $\theta=\pi\,.$
Therefore it is desirable to have two free parameters for $\tilde{f}\,.$
Also since $f_{a}$ is an even function,
$\tilde{f}$\ is desired to be an even function.
We also impose the condition that
the corresponding function $h$
in eq.~(\ref{e23}) has an analytic expression.
The simplest choice satisfying these conditions
is the following $\tilde{f}_{\alpha,\beta}\,;$
\begin{equation}
\tilde{f}_{\alpha,\beta}(\theta)=
\frac{\tilde{N}_{\alpha,\beta}}
 {2\cosh(\alpha\theta)+2\beta}
\,,\ \alpha>0\,,\ \beta>-1\,,
\end{equation}
where $\tilde{N}_{\alpha,\beta}$ is a normalization constant;
\begin{equation}
\label{e35}
\tilde{N}_{\alpha,\beta}=
  \left\{
  \begin{array}{ll} \displaystyle
  \alpha \frac{\sqrt{\beta^2-1}}
   {2 \tanh^{-1} (A\,B)} \,,\
  &
  \beta>1\,,
  \\
  \displaystyle
  \alpha \frac{1}{A} \,,\
  &
  \beta=1\,,
  \\
  \displaystyle
  \alpha \frac{\sqrt{1-\beta^2}}
   {2 \tan^{-1} (A\,B)} \,,\
  &
  -1<\beta<1\,,
  \end{array}
  \right.
\end{equation}
where we put $\displaystyle A=\tanh \frac {\pi\alpha}{2}\,,$
and $\displaystyle B=\sqrt{\frac{|\beta-1|}{\beta+1}}\,.$
The corresponding function $h$ in eq.~(\ref{e23}) is;
\begin{equation}
\label{e31}
h_{\alpha,\beta}(x)=
  \left\{
  \begin{array}{ll}
  \hbox{\vrule height17pt depth7pt width0pt}
  \displaystyle
  {2\over \alpha}
  \tanh^{-1}
  \left(
  B^{-1}
   \tanh((2x-1)\tanh^{-1}(A\,B))
  \right) \,,\
  &
  \beta>1\,,
  \\
  \hbox{\vrule height17pt depth7pt width0pt}
  \displaystyle
  {2\over \alpha}
  \tanh^{-1}
  \left(
   (2x-1) A
  \right) \,,\
  &
  \beta=1\,,
  \\
  \hbox{\vrule height17pt depth7pt width0pt}
  \displaystyle
  {2\over \alpha}
  \tanh^{-1}
  \left(
  B^{-1}
   \tan((2x-1)\tan^{-1}(A\,B))
  \right) \,,\
  &
  -1<\beta<1\,,
  \end{array}
  \right.
\end{equation}
where $A$ and $B$ are as above.

The next step is to choose
$\alpha=\alpha(a)$ and $\beta=\beta(a)$ as functions of $a\,.$
In principle, they should be chosen so as to
optimize the acceptance rate $R=R(a)\,.$
Here, we search for a solution that satisfies
a condition that the minimum in the definition of $R(a)$
(i.e. in the right hand side of eq.~(\ref{e25}))
is achieved at $\theta=0\,.$
We impose this condition to avoid ^^ if-branches' in the
resulting computer program.
If the minimum is attained at different values of $\theta\,,$
we will need if-branches according to the different values of $\theta\,.$

We have an argument that the optimal solution under this condition,
which we shall refer to as the
^^ optimized cosh' method, is given by choosing
$\alpha=\alpha(a)$ and $\beta=\beta(a)$ in  eq.~(\ref{e31})
to satisfy the following:

\begin{equation}
\label{e32}
\alpha(a)=\sqrt{3a-1}\,,\ \beta(a)=2-\frac{1}{a}\,,
\ \ \mbox{if}\
a \geq a^{o}\,,
\end{equation}
\begin{equation}
\label{e33}
\frac{\cosh(\pi\alpha(a))-1}{\alpha(a)^{2}} = \frac{\exp(2a)-1}{a}\,,\
\beta(a)=\frac{\alpha(a)^{2}}{a}-1\,,
\ \ \mbox{if}\
a^{o}>a \geq a^{*}\,.
\end{equation}
Here $a^{o}$ and $a^{*}$ are positive constants
satisfying $a^{o}>a^{*}$, uniquely determined by,
\begin{eqnarray}
\label{e38}
\frac{\exp(2a^o)-1}{a^o} &=& \frac{\cosh(\pi \sqrt{3a^o-1})-1}{3a^o-1}\,,\
\\
\label{e36}
\frac{\exp(2a^*)-1}{a^*} &=& \frac{\pi^2}{2}\,.
\end{eqnarray}
Their numerical values are $a^{*} \approx 0.799$
and $a^{o} \approx 5.04$.
The function $g=g_a$ in eq.~(\ref{e24}) is , for $a \geq a^{*}\,,$
\begin{equation}
\label{e37}
g_a(x)=\exp(-a\,G_a (h_{\alpha(a),\beta(a)}(x)))
\end{equation}
with
\begin{equation}
G_a (\theta)=1-\cos \theta - \frac{1}{a} \log
\left( 1+ \frac{1}{1+\beta(a)} (\cosh(\alpha(a) \, \theta)-1) \right) \,.
\end{equation}

For the parameter range of $0<a<a^{*}\,,$
we have to take a limit
$\alpha \downarrow 0$ with
$\displaystyle \frac{\alpha(a)^2}{1+\beta(a)}$ fixed to
$2\,\pi^{-2}(\exp(2a)-1)\,.$
We have, in place of
eq.~(\ref{e31}),
\begin{equation}
\label{e34}
h_{\gamma}(x)=
   \frac{1}{\gamma}
   \tan((2x-1)\tan^{-1}(\pi\,\gamma)) \,,\
\end{equation}
where $\gamma = \gamma(a)$ is
\begin{equation}
\gamma(a)=\pi^{-1}\sqrt{\exp(2a)-1} \,,\ \ \mbox{if}\ 0<a<a^{*}\,.
\end{equation}
The function $g=g_a$ in eq.~(\ref{e24}) is
\begin{equation}
g_a(x)=\exp(-a\,G_a (h_{\gamma(a)}(x)))
\end{equation}
with
\begin{equation}
G_a (\theta)=1-\cos \theta - \frac{1}{a} \log
\left( 1+ \gamma(a)^2 \theta^2 \right) \,.
\end{equation}
The distribution is reduced to the Cauchy distribution:
\begin{equation}
\tilde{f}_{\gamma}(\theta)=
\frac{\tilde{N}_{\gamma}}
{1+ \gamma^2 \theta^2 }\,,
\end{equation}
where
\[
\tilde{N}_{\gamma}=
         \frac{\gamma}
   {2 \tan^{-1} (\pi \gamma)} \,.
\]

See appendix~\ref{sb} for
the proof that these formulae correctly generate
a random $U(1)$ variable, and
arguments for our choice of the parameters.

The acceptance rate $R=R(a)$ for the ^^ optimized cosh' method is
given by
\begin{eqnarray*}
R(a) &=
\displaystyle
         \frac{\tilde{N}_{\alpha(a),\beta(a)}}
              {2\, N_a \exp(a) (1+\beta(a)) }\,,
& a \geq a^{*}\,,
\\
     &=
\displaystyle
         \frac{\gamma(a)}
              {2\, N_a \exp(a) \tan^{-1} (\pi \gamma(a))} \,,
& a^{*}> a> 0\,,
\end{eqnarray*}
where $N_a$ and $\tilde{N}_{\alpha,\beta}$ are defined in
eq.~(\ref{e27}) and eq.~(\ref{e35}), respectively.
Note the high acceptance rate for both small $a$ and large $a$:
\[
\begin{array}{ccll}
R(a) & \approx &
\displaystyle
1-\frac{1}{3}a+\frac{71}{180}a^2 \,,
& a \ll 1 \,,
\\
R(a) & \to &
\displaystyle
\frac{\sqrt{2 \pi}}{2 \log(2+\sqrt{3})} \approx 0.95,
& a \to \infty \,.
\end{array}
\]
See Fig.~1 for the acceptance rate for full range of $a$.
The acceptance rate for the ^^ optimized cosh' method keeps
more than 90\% for all values of $a$,
including the ^^ zero temperature' limit $a\rightarrow\infty\,.$

\section{Proposed algorithm.}  \label{s4}
%section 4
The acceptance rate for the ^^ optimized cosh' method is high, but
to obtain the parameter $\alpha(a)$ for $a^{*} \leq a < a^{o}\,,$
one has to solve a transcendent equation eq.~(\ref{e33}).
Also, one has to use different formulae for
$0<a<a^{*}\,,\ a^{*} \leq a < a^{o}\,,$ and $a^{o} < a\,,$
which will cause ^^ if-branches'
that will considerably lower the efficiency
when using with vectorized processors.

One may, for example, use Newton method to solve equations numerically,
but here instead we approximate the function
$\alpha(a)$ in eq.~(\ref{e33}) directly by a function
which is explicitly expressible on computer programs
without using if-branches for all $a$, and designed to keep high
acceptance rate.
The if-branches are avoided in such a way that we have
$-1< \beta <1$ for all $a \in (0,\infty)$.

We shall give an example of realistic choice.

\bigskip

\begin{enumerate}

\item
Define $\alpha(a)$ and $\beta(a)$ by,
\begin{eqnarray*}
\alpha(a) &=& \min\{\sqrt{a \, (2-\epsilon)}\,,\
\max\{\sqrt{\epsilon a}\,,\ \delta(a)\}\}\,,
\\
\beta(a) &=& \max\{
\frac{\alpha(a)^{2}}{a} \,,\
\frac{\cosh(\pi\alpha(a))-1}{\exp(2a)-1} \} -1 \,,
\end{eqnarray*}
where
\[
\delta(a)=0.35\,\max\{0\,,\ a-a^{*}\}+1.03\,\sqrt{\max\{0\,,\ a-a^{*}\}}\,,
\]
and $ \epsilon=0.001 $.
$a^*=0.798953686083986$ is the constant defined by eq.~(\ref{e36}).

\item
Define functions $h_{\alpha,\beta}$ and $g_a$ by
\[%\begin{equation}
h_{\alpha,\beta}(x)=
  \frac{2}{\alpha} \tanh^{-1}
  \left(
   \sqrt{ \frac{1+\beta}{1-\beta} }
   \tan( (2x-1) \tan^{-1}(
    \sqrt{\frac{1-\beta}{1+\beta}} \, (\tanh \frac {\pi\alpha}{2})
                          )
       )
  \right) \,,\
\]%\end{equation}
and
\[%\begin{equation}
g_a(x)=\exp(-a\,G_a (h_{\alpha(a),\beta(a)}(x)))
\]%\end{equation}
with
\[%\begin{equation}
G_a (\theta)=1-\cos \theta - \frac{1}{a} \log
\left( 1+ \frac{1}{1+\beta(a)} (\cosh(\alpha(a) \, \theta)-1) \right) \,.
\]%\end{equation}

\item
Let $\omega_{j}$ and $\omega'_{j}$ with $j=1,2,3,\cdots,$
be two sequences of independent random variables uniformly
distributing in $[0,1)$.
Define a subsequence
\[ \tilde{\omega}_{i}=\omega_{j_{i}}\,,\ i=1,2,3,\cdots\,, \]
of the sequence $\{\omega_{j}\}$
by selecting the numbers $j=j_{i}$ that satisfy
\[ \omega'_{j}\leq g_a(\omega_{j})\,. \]

The sequence
\[%\begin{equation}
h_{\alpha(a),\beta(a)}(\tilde{\omega}_{1})\,,\
h_{\alpha(a),\beta(a)}(\tilde{\omega}_{2})\,,\
h_{\alpha(a),\beta(a)}(\tilde{\omega}_{3})\,,\
\cdots,
\]%\end{equation}
is the random $U(1)$ variable, a sequence whose distribution is
\[
P([\theta,\theta+d\theta])=N_{a} \exp(a \cos \theta)\,d\theta
\,,\ -\pi\leq \theta<\pi\,,\ a> 0\,,
\]
\[
\frac{1}{N_{a}} = \int_{-\pi}^{\pi} \exp(a \cos\theta)\,d\theta\,.
\]

\end{enumerate}

\bigskip

We will refer to this choice as the ^^ proposed cosh' method.
One can explicitly check that the choice of parameters satisfies
$-1< \beta(a) <1$ for all $a \in (0,\infty)$.
The functions $h_{\alpha,\beta}$ and
$g_a$ therefore are obtained from section~\ref{s3}.
One can see that this choice of parameters
satisfies the conditions (\ref{eb2}), (\ref{eb3}), and (\ref{eb4}),
which implies that the ^^ proposed cosh' method
correctly generates $U(1)$ random variables.
The acceptance rate $R(a)$ for ^^ proposed cosh' method is
given in Figure~1.
Note that the acceptance rate is high for all values of $a$.
The minimum of $R(a)$ for ^^ proposed cosh' method is
$R(a=\infty) \approx 0.88$.

\section{Efficiency test.}  \label{s5}
%section 5
We will show our results of efficiency test for ^^ propsed cosh' method.

In Monte Carlo simulations,
we try to generate a number in the random $U(1)$ variable,
for each trial update of a site or link variable.
As explained in section~\ref{s2}, this is to generate
two independent uniform random number $\omega_j$ and
$\omega'_j$ and decide whether to accept or reject
$h(\omega_j)$.
We may alternatively decide to try at most $n$-times,
namely to try
$\omega_j,\ \omega_{j+1},\ \cdots,\ \omega_{j+n-1},$
before deciding that this site variable was not updated.
(We shall call this number $n$, the iteration number,
and call this one set of trial, an update.)
This will effectively improve
the acceptance rate to
\begin{equation}
R_{n}=1-(1-R)^{n}\,,
\end{equation}
where $R$ is the original acceptance rate.
On the other hand, this will increase
the time consumption per update by an average rate of roughly
$R^{-1}$ when $n(1-R)^n \ll 1\,.$

In other words, one must consider the possibility of
iterating an algorithm with low acceptance rate but high speed,
such as iterating the ^^ direct' method.
Improvement of the efficiency resides in
a balance of high acceptance rate
and simple (fast) program.

Efficiency is a measure of
average speed of producing
random $U(1)$ variable.
 {}From the above consideration on Monte Carlo updates,
we may define the efficiency
by the average time consumption for an update
with iteration number $n$ chosen so that the effective
acceptance rate $R_n$ exceeds some fixed rate, say,
\begin{equation}
R_n > 0.9\,.
\end{equation}
This definition is useful
when considering Monte Carlo calculations.

We note that this is not the only possible definition of efficiency.
Alternatively, one may use the criteria to iterate
until a variable is updated.  In appendix~A~of~\cite{EGS} this
criteria is adopted to measure the speed of their methods.
However, except for the discreteness of iteration number
(and modulo multiplication of a constant),
the two criteria essentially measure the same quantity,
so the conclusion will not change.

The best possible choice depends on the machine to be used.
We checked our choice in section~\ref{s4} by
HITACH S820 in Computer Centre of University of Tokyo,
which uses a pipelined vector processor.
We measured the efficiency
of generating random $U(1)$ variable
defined as above,
with average taken over $4 \times 10^6$ updates.
The ^^ proposed cosh' method (without iteration) has acceptance rate of
more than 90\% for $a\leq8\,.$
We compared the speed of ^^ proposed cosh' method
with the iterated ^^ direct' method for $a\leq8\,,$
with the condition that the number of iteration is chosen
to keep the acceptance rate to be more than 90\%.

The measured acceptance rates were in good agreement
with the theoretical predictions in section~\ref{s3}
and in Figure~1 (within 0.1\% acuracy).

The results of the efficiency test is given in
Figure~2, which
shows that the ^^ proposed cosh' method is efficient
for all values of $a$, especially for large $a$.
Note that for the ^^ direct' method, the CPU time increases as
$a$ increases, which is due to the decrease in the acceptance rate $R(a)\,.$
Since $R(a) \to 0$ as $a \to \infty\,,$ the CPU time for
the ^^ direct' method will tend to $\infty$\ as $a \to \infty\,.$
The acceptance rate for the ^^ proposed cosh' method is
greater than $0.88647$ for all values of $a$,
consequently the CPU time is bounded for all $a$, and furthermore,
the method is suitable for vector processors.

\section{Concluding remarks.}  \label{s6}
%section 6
We have shown a simple method for
generating random $U(1)$ variables.
Our method, based on rejection algorithm, achieves
high acceptance rate and low time consumption
for all values of the parameter $a$,
including the ^^ low temparature limit', $a \to \infty$ (section~\ref{s5}).
The only requirement
on the hardware for our method to be efficient, is that
the computer is quick in calculating elementary functions
such as $\exp(x),\ \tan(x),$ and their inverse functions.
Many present day computers
which are equipped with co-processors for floating point calculations
are suitable for our methods.
Our method which keeps uniformly high acceptance rate for all values of $a$,
is particularly suitable to parallel or pipeline vector processors.

 {}From the general consideration given in section~\ref{s3},
the use of the ^^ cosh' distribution and the Cauchy distribution
which we proposed as approximate distributions,
will be efficient for generating random variables
taking values in a finite interval (e.g. $[-\pi,\pi]$) and
whose distribution $f_a(\theta)$
is an even function and takes maximum at $\theta=0$, and
behaves like $f_a(\theta) \approx const.-a \theta^2$ near $\theta=0\,,$
with a parameter $a>0$ that controls the sharpness of the peak.
This is a common feature of weight functions
for statistical mechanical systems
with one component spins and link variables.

We would like to mention a couple of methods which appears in the literature,
both of which are based on rejection methods, but use different
approximate distributions $\tilde{f}\,.$
The approximate distribution adopted in \cite{Mo} is
\[
\tilde{f}(\theta)=\tilde{N}_{a} \exp(a\,(1-\frac{2}{\pi}|\theta|))\,,
\]
where $\tilde{N}_a$ is a normalization constant.
The acceptance rate $R(a)$ for this choice is, from eq.~(\ref{e27}),
\[
R(a)=\frac{1}{N_a \exp(a)} \frac{a \exp(-c\,a)}{\pi\,(1-\exp(-2a))}\,,
\]
where $N_a$ is as in eq.~(\ref{e27}) and
$\displaystyle c=
\frac{2}{\pi}\,\sin^{-1}(\frac{2}{\pi})+\sqrt{1-(\frac{2}{\pi})^2}-1
\approx 0.2105$ is a positive constant.
In particular,
at ^^ low temperature' $a \gg 1\,,$
\[
R(a) \approx \sqrt{\frac{2a}{\pi}} \exp(-c\,a) \,,
\]
which rapidly approaches $0$ as $a \to \infty\,.$
Therefore this choice suffers from the same problem as with
^^ direct' method that as $a \to \infty\,,$ the CPU-time
to keep 90\% effective acceptance rate tends to $\infty\,.$
In fact, we have performed an efficiency test as explained in
section~\ref{s5} for this choice of approximate distribution,
and found that for $a>3\,,$ the ^^ proposed cosh' method
is faster.

The approximate distribution of
\cite{EGS} is the Gaussian distribution which, in our notation, is
\[
\tilde{f}(\theta)=\tilde{N}_{a} \exp(-\alpha(a)\,\theta^2)\,,
\]
where
\[
\frac{1}{\tilde{N}_{a}}
= \int_{-\infty}^{\infty} \exp(-\alpha(a)\,\theta^2)\,d\theta\,,
\]
is a normalization constant,
and $\displaystyle \alpha(a)=\frac{2}{\pi^2} \max(a,\frac{1}{4})\,.$
(To be precise, they adopt ^^ direct' method for small $a$ and
Gaussian distribution for large $a\,.$
We focus our attension on the large $a$ case where the
simple ^^ direct' method is not effective.)
\cite{EGS} quotes \cite{K} for an algorithm of generating
the Gaussian random variable.

The original distribution $f_a(\theta)$ is now considered
as a distributionon $\theta \in {\bf R}\,,$ where
$f_a(\theta)=0$ if $ |\theta| > \pi\,.$
The formula similar to eq.~(\ref{e25})
(with $-\pi \le \theta < \pi$ replaced by
$\theta \in {\bf R}$) holds in the present case, and
the acceptance rate $R(a)$ for the Gaussian distribution is,
\[
R(a)=\frac{1}{N_a \exp(a)} \sqrt{\frac{2a}{\pi^3}}\,,
\]
where $N_a$ is as in eq.~(\ref{e27}).  This coincides with the
results in \cite{EGS}.
At $a \to \infty\,,$ $\displaystyle R(a) \to \frac{2}{\pi} \approx 0.6366\,.$
Since $R(a)$ does not tend to zero, the effective CPU-time consumption
is bounded for all range of $a$, which is of course a nice feature shared
with our methods.
The acceptance rate of ^^ proposed cosh' method is considerably
higher than that of the Gaussian method for all values of $a\,.$
As for a quantitative comparison of CPU-time consumption,
\cite{EGS} compares
the CPU-time between the ^^ direct' method and the Gaussian method,
and finds that (with their machine) for $a>1.5$ the Gaussian method
becomes faster than the ^^ direct' method, and slows down somewhat
as $a$ is increased further.
(Roughly 12\% decrease in speed from $a=1.5$ to $a=\infty\,.$)
As can be seen from
Figure~2,
at $a=1.5$ the ^^ proposed cosh' method is already faster
(the ratio in speed is roughly 1.6)
than the ^^ direct' method, and it practically does not slow down
if $a$ is increased.
(A rough estimate shows that only 3\% decrease in speed occurs
{}from $a=1.5$ to $a=\infty\,.$)
Therefore we may conclude that the ^^ proposed cosh' method is faster than
the Gaussian method for $a>1.5\,.$

\bigskip

\bigskip

\noindent
{\bf Acknowledgements.} \nopagebreak
\\[1em]
We would like to thank Prof. Y. Oyanagi for very helpful discussions
and encouragements.
We would also like to thank Prof. A. D. Sokal for
instructive comments and bringing basic references to
our attention.

\bigskip

\bigskip

\appendix
\section{}  \label{sa}
%appendix a
In this Appendix, we prove the statements in section~\ref{s2}.

Let $p$ be the density function for the distribution
of the random variable (\ref{e26}):
\[
p(\theta)\,d\theta=Prob[\, h(\tilde{\omega}_{i})\in [\theta,\theta+d\theta] \,]
\,,\]
where
\[
Prob[\, P(i) \,] \equiv
\lim_{N\rightarrow\infty} \frac{1}{N}
\sharp \{ 1\leq i\leq N\,|\ P(i) \}\,.
\]
The statement $p=f_{a}$ is proved as follows.
By eq.~(\ref{e23}) we have
\[
h^{-1}(\theta+d\theta)=h^{-1}(\theta)+\frac{1}{h'(h^{-1}(\theta))}\,d\theta
=h^{-1}(\theta)+\tilde{f}(\theta)\,d\theta\,,
\]
hence
\begin{equation}
p(\theta)\,d\theta = Prob[\, \tilde{\omega}_{i}\in
[h^{-1}(\theta),h^{-1}(\theta)+\tilde{f}(\theta)\,d\theta] \,]\,.
\label{ea1}
\end{equation}

Fix $0\leq x<1\,.$
By definition we have $0\leq g(x) \leq 1\,.$
Since $\omega'_{j}\,,\ j=1,2,3,\cdots,$ is a uniform random variable,
we see that
the conditional acceptance rate for $\omega_{i}$
under the condition that $\omega_{i} \in [x,x+dx]\,,$
is $g(x)\,.$
Therefore,
\[
Prob[\,\tilde{\omega}_{i} \in [x,x+dx]\,]
= Prob[\, \omega_{i} \in [x,x+dx]\,] \times g(x)
\propto g(x)\,dx\,,
\]
where we also used the fact that
$\omega_{i}\,,\ i=1,2,3,\cdots,$ is a uniform random variable.
Insert this equation in eq.~(\ref{ea1}) and use eq.~(\ref{e24}) to obtain
\[
p(\theta)\,d\theta \propto g(h^{-1}(\theta))\,\tilde{f}(\theta)\,d\theta
\propto f_{a}(\theta)\,d\theta\,.
\]
Since $f_{a}$ is correctly normalized, we have $p(\theta)=f_{a}(\theta)\,.$
This completes the proof.
\vspace{1ex}

The proof that the acceptance rate is equal to $R$ in eq.~(\ref{e25})
is as follows.
For each $x$ with $0 \leq x<1\,,$
the conditional acceptance rate for $\omega_{i}$
under the condition that $\omega_{i} \in [x,x+dx]\,,$
is $g(x)\,.$
Therefore
\begin{eqnarray*}
\mbox{The acceptance rate} &=& \int_{0}^{1} g(x)\,dx \\
&=& R\, \int_{0}^{1} \frac{f_{a}(h(x))}{\tilde{f}(h(x))}\,dx \\
&=& R\, \int_{-\pi}^{\pi} \frac{f_{a}(\theta)}{\tilde{f}(\theta)}\,
\tilde{f}(\theta)\,d\theta = R\,.
\end{eqnarray*}
This completes the proof.

\section{}  \label{sb}
%appendix b
In this Appendix,
we give a proof that the ^^ optimal cosh' method
in section~\ref{s3} correctly gives the random $U(1)$-variables,
and also
we give the argument for the choice of the parameters.

We consider the case $a > a^*$.
The proof for the case $0<a \le a^*$ is similar.

By explicit calculation,
one sees that
$h_{\alpha,\beta}$ of eq.~(\ref{e31}) satisfies
eq.~(\ref{e23}).
Therefore it suffices to show that eq.~(\ref{e37})
satisfies eq.~(\ref{e24}) with eq.~(\ref{e25}).

Define,
\begin{eqnarray}
G(\theta) &=& \frac{1}{a} \log \left\{
            \frac{\tilde{f}(\theta)}{f_a(\theta)}
            \frac{f_a(0)}{\tilde{f}(0)} \right\}
\\
          &=& 1-\cos \theta - \frac{1}{a} \log
\left( \frac{1}{1+\beta} (\cosh(\alpha \, \theta)+\beta) \right) \,.
\label{eb1}
\end{eqnarray}
($G$ depends on three free parameters $a$, $\alpha$, and $\beta$.
We suppress the parameter dependences for the moment.)
Then eq.~(\ref{e24}) and eq.~(\ref{e25}) imply
\begin{equation}
\label{eb10}
g(x) = \exp \{ -a\, (G(h(x)) - \min_{\theta \in [-\pi,\pi]} G(\theta)) \}\,,
\end{equation}
and
\[
R(a) = \frac{\tilde{f}(0)}{N_a \exp(a)}
       \exp \{ a \min_{\theta \in [-\pi,\pi]} G(\theta) \}\,.
\]

Comparing eq.~(\ref{eb1}) with eq.~(\ref{e37}),
one sees that the results in section~\ref{s3} are correct
if
\begin{equation}
\label{eb5}
\min_{\theta \in [-\pi,\pi]} G(\theta)=0 \,.
\end{equation}

\medskip
{\bf Proposition.}
Let $a>0\,,\ \alpha>0\,,\ \mbox{and}\ \beta>-1\,.$
If $G$ satisfies the three conditions
\begin{equation}
\label{eb2}
G''(0) = 1-\frac{1}{a} \frac{\alpha^2}{1+\beta} \ge 0 \,,
\end{equation}
\begin{equation}
\label{eb3}
G^{(4)}(0)
= -1-\frac{1}{a} \frac{\alpha^4(\beta-2)}{(1+\beta)^2} \ge 0 \,,
\end{equation}
\begin{equation}
\label{eb4}
G(\pi) = 2-\frac{1}{a} \log
\left( \frac{\cosh(\pi\alpha)+\beta}{1+\beta} \right) \ge 0 \,,
\end{equation}
then $G$ satisfies
\begin{equation}
\label{eb13}
G(x) \ge 0 \,, \ \ -\pi \le x \le \pi \,.
\end{equation}

Assume for the moment that this Proposition is true.
It is easy to see by explicit calculations
that $\alpha=\alpha(a)$ and $\beta=\beta(a)$
defined by eq.~(\ref{e32}) or eq.~(\ref{e33})
satisfy the conditions (\ref{eb2}), (\ref{eb3}), and (\ref{eb4}),
and $\alpha(a)>0$ and $\beta(a)>-1\,,$
for all $a>a^* \,.$
Since
\[
G(0)=0\,,
\]
the Proposition implies that eq.~(\ref{eb5}) is satisfied
for all $a>a^* \,.$

It remains to prove the Proposition.

\medskip
{\bf Proof of the Proposition.}
Since $G$ is an even function, it is sufficient to prove
$G(x) \ge 0 $ for $0 \le x \le \pi \,.$

 {}From (\ref{eb2}) and (\ref{eb3}) it follows that
$\alpha^2 (2-\beta) \ge \beta+1 \,.$
The equality holds if and only if
$a (\beta+1) = \alpha^2$ and $\alpha^2 = 3 a - 1\,,$
which, with (\ref{eb4}) implies
$(3a-1)(\exp(2a)-1) \ge a (\cosh(\pi \sqrt{3a-1})-1)\,.$
This is equivalent to $a>a^o \, (>3/2)\,,$
where $a^o$ is defined by eq.~(\ref{e38}).
Therefore, if we define a set $D$ by
\begin{eqnarray*}
D & = & D_1 \cup D_2 \,,
\\
D_1 & \equiv &
\{(a,\alpha,\beta) \in (0,\infty)^2 \times (-1,\infty) \,|\
\alpha^2 (2-\beta) > \beta+1  \ge \alpha^2 /a \} \,,
\\
D_2 & \equiv &
\{(a,\alpha,\beta) \in (0,\infty)^2 \times (-1,\infty) \,|\
\alpha^2 = 3a-1 \,,\ \beta+1=\alpha^2/a \,,\ a>3/2 \}\,,
\end{eqnarray*}
it is sufficent to prove that for all $(a,\alpha,\beta) \in D$ and
$0 \le x \le \pi\,,$ (\ref{eb4}) implies $G(x) \ge 0\,.$

\medskip \noindent {\bf Step 1.}
Fix $(a,\alpha,\beta) \in D\,.$
Put $g(x) \equiv G'(x)\,.$
(This $g$ has nothing to do with $g$ in eq.~(\ref{eb10}).)
Then we have
\begin{eqnarray}
\label{eb6}
f(x) & \equiv & g(x)+g''(x)
\\
\nonumber
& = & \frac{\alpha^5 \sinh(\alpha x)}{a (\beta+ \cosh(\alpha x))^3}
\, h(\alpha^{-2} (\cosh(\alpha x)-1)) \,,
\end{eqnarray}
where
\[
h(y) \equiv -y^2 + (\beta - 2 \alpha^{-2} (\beta+1)) y
+ \alpha^{-4} (\beta+1) (\alpha^2 (2-\beta)-(\beta+1)) \,.
\]
Since $(a,\alpha,\beta) \in D \,,$ we see that
there exists one and only one
positive root $y=y_0$ of $h(y)=0$ and that
\begin{eqnarray*}
h(y)>0 \,,\ & \mbox{if}\ & 0<y<y_0 \,,
\\
h(y)<0 \,,\ & \mbox{if}\ & y>y_0 \,.
\end{eqnarray*}
Therefore if we let $x=x_0$ to be the unique positive solution
to the equation $ \alpha^{-2} (\cosh(\alpha x)-1) = y_0 \,,$
we have
\begin{eqnarray}
\label{eb7}
f(x)>0 \,,\ & \mbox{if}\ & 0<x<x_0 \,,
\\
\label{eb8}
f(x)<0 \,,\ & \mbox{if}\ & x>x_0 \,.
\end{eqnarray}

Note that $g(0)=0$ and $g'(0) \ge 0$ if $(a,\alpha,\beta) \in D \,.$
 {}From eq.~(\ref{eb6}) and eq.~(\ref{eb7}) we therefore conclude
\begin{equation}
\label{eb9}
g(x) > 0 \,,\ \mbox{if}\ 0<x \le x_0 \ \mbox{and}\ 0<x<\pi \,.
\end{equation}
(The conclusion may be easily understood if one notes that
eq.~(\ref{eb6}) is an equation of motion of harmonic oscillation
with external force $f\,.$)

The equations (\ref{eb6}), (\ref{eb8}), (\ref{eb9}), imply
\begin{eqnarray}
\label{eb11}
G'(x)=g(x)>0 \,,\ &
\mbox{if}\ & 0<x \le x_0 \ \mbox{and}\ 0<x<\pi \,.
\\
\label{eb12}
g(x)+g''(x)<0 \,,\ & \mbox{if}\ & x>x_0 \,.
\end{eqnarray}

\medskip \noindent {\bf Step 2.}
Fix $\alpha>0$ and $t \equiv a \alpha^{-2} (\beta+1) \ge 1\,,$
and let $a$ vary with the restriction $(a,\alpha,\beta) \in D\,.$
The allowed region of $a$ differs by the values of $t$ and $\alpha$:
\begin{enumerate}
\item
$t>1\,,$ or $t=1$ and $\alpha \le \sqrt{7/2}\,.$
\\
In this case, $(a,\alpha,\beta) \in D$ is equivalent to
$a>(\alpha^2+1)/3\,.$
\item
$t=1$ and $\alpha > \sqrt{7/2}\,.$
\\
In this case, $(a,\alpha,\beta) \in D$ is equivalent to
$a \ge (\alpha^2+1)/3\,.$
\end{enumerate}

Note that $g(x)=g_a(x)$ is continuous
(uniformly continuous on compact sets in $(0,\pi]$ w.r.t. $x$)
and increasing in $a\,,$
and $x_0=x_0(a)$ is continuous in $a\,.$
Also,
\[
\lim_{a \to \infty} g(x) = \sin x
\]
uniformly on compact sets in $(0,\pi]\,.$

We claim that for every $a$ (such that $(a,\alpha,\beta) \in D$),
and for any $x_1$ and $x_3$ satisfying
$g_a(x_1)>0\,,\ g_a(x_3)>0\,,$ and $0<x_1<x_3 \le \pi\,,$
we have $g_a(x) > 0 \,,\ x \in [x_1,x_3]\,.$

Assume this is wrong:
Assume that for $a=a_0$ and $0<x_1<x_2<x_3 \le \pi$
we have $g_{a_0}(x_1)>M\,,\ g_{a_0}(x_2) \le 0\,,$ and
$g_{a_0}(x_3)>M\,,$
where $M$ is a positive constant.
Since $g_a(x)$ is increasing in $a\,,$ we have
\[
g_a(x_1)>M\,,\ g_a(x_3) > M \,,\ a \ge a_0 \,.
\]
Put
\[
q(a) \equiv \min_{x_1 \le x\le x_3} g_a(x) \,.
\]
Then $q(a)$ is continuous in $a$ and
$\displaystyle \lim_{a \to \infty} q(a) > 0 \,.$
Therefore there exists $a_1 \ge a_0$ such that
$q(a_1)=0 \,,$ which further implies that
$g_{a_1}(x) \ge 0 \,,\ x_1 \le x \le x_3 \,,$ and that
there exists $x_4$
satisfying $x_1 < x_4 < x_3$ and $g_{a_1}(x_4)=0 \,.$
In particular, $g_{a_1}(x_4)=0$ and $g''_{a_1}(x_4) \ge 0$ hold,
which contradicts eq.~(\ref{eb11}) and eq.~(\ref{eb12}).
Hence the claim is proved.

\medskip \noindent {\bf Step 3.}
Fix $(a,\alpha,\beta) \in D\,.$
 {}From the claim and eq.~(\ref{eb11}) we see that
either $g(x)=G'(x) >0$ for $0<x<\pi\,,$
or there exists $x'$ such that $0<x'<\pi\,,$ and
$g(x)>0$ for $0<x<x'\,,$ and $g(x)<0$ for $x'<x \le \pi\,.$
Hence $G(x)$ is either increasing in $0 <x< \pi$ or
has just one peak and no valley.
Since $G(0)=0$ and $G(\pi) \ge 0\,,$
we have $G(x) > 0 \,,\ 0<x<\pi \,.$
This completes the proof.

\bigskip

We now turn to the argument for the choice of the parameters.
We want to choose the parameters so that the acceptance rate
is as large as possible.
As stated at the end of section~\ref{s2},
it is better to have as flat $g(x)$ as possible, hence
a flat $G(x)\,.$  Thus we require $G''(0)=0\,.$
We impose the condition that the minimum of $R(a)$ is
achieved at $\theta=0\,.$ Note that this is equivalent to
assuming eq.~(\ref{eb13}).
As a necessary condition, we have
$G^{(4)}(0) \ge 0$ and $G(\pi) \ge 0 \,.$
(By the Proposition, we know that these are sufficient to
ensure eq.~(\ref{eb13}).)
As we want to have flat $G(x)\,,$
it should be best to have
either $G^{(4)}(0) = 0$ or $G(\pi) = 0 \,.$
If one draws a graph of these three conditions,
in $(\alpha,\beta)$-plane,
one easily sees that the choice given in the section~\ref{s3}
is the one that we are looking for.

\section{}  \label{sd}
%appendix d
In this Appendix, we give a sample FORTRAN program for generating
random $U(1)$ variables, using ^^ proposed cosh' method.
When the subroutine U1RND is called with the parameter $a$ as
the first argument, it returns a random $U(1)$ variable
as the second argument.

Note that this sample program is different
{}from the program used to test the efficiency disscussed in
section \ref{s5}, where we modified the program in favor of
a pipelined vector processor.

\bigskip

% \documentstyle{article}
% \begin{document}
\begin{verbatim}
      SUBROUTINE U1RND(A,G)
C VARIABLES FOR U(1) RANDOM NUMBERS
      REAL PI,A,AS,EPS,DEL,ALP,BET,DAP,DL1,DL2,BT1,H,G,GH,H1
      PARAMETER (PI=SNGL(3.14159265358979D0))
      PARAMETER (AS=SNGL(0.798953686083986D0))
      PARAMETER (EPS=0.001, DL1=0.35, DL2=1.03)
C VARIABLES FOR UNIFORM RANDOM NUMBERS
      INTEGER C30,CRND,IRND
      REAL RND
      REAL*8 C31
      PARAMETER (C30=2**30, C31=2D0**31, CRND=48828125)
      DATA IRND/1000001/
C
      DAP=MAX(0,A-AS)
      DEL=DL1*DAP+DL2*SQRT(DAP)
      ALP=MIN(SQRT(A*(2-EPS)), MAX(SQRT(EPS*A),DEL))
      BET=MAX(ALP*ALP/A, (COSH(PI*ALP)-1)/(EXP(2*A)-1))-1
      BT1=SQRT((1+BET)/(1-BET))
C
 1    CONTINUE
C
      R=IRND/C31
      IRND=IRND*CRND
      IF (IRND .LT. 0) IRND=(IRND+C30)+C30
C
      H1=BT1*TAN((2*R-1)*ATAN(TANH(PI*ALP/2)/BT1))
      H=ALOG((1+H1)/(1-H1))/ALP
      G=EXP(-A*(1-COS(H)))*(COSH(ALP*H)+BET)/(1+BET)
C
      R=IRND/C31
      IRND=IRND*CRND
      IF (IRND .LT. 0) IRND=(IRND+C30)+C30
C
      IF (G .LT. R) GOTO 1
      RETURN
      END
\end{verbatim}
% \end{document}

\bigskip

\bigskip

\newpage
%figure captions
\noindent
{\bf \large Figure captions.} \nopagebreak
\\[1em]
Figure 1:
Acceptance rate $R(a)$;  ^^ direct' method (lowest curve),
^^ optimized cosh' method (highest curve),
and ^^ proposed cosh' method(curve in between).
\\[1em]
Figure 2:
Time consumption $T$ to keep 90\% acceptance rate $R(a)>0.9\,.$
Iterated ^^ direct' method ($+$) and ^^ proposed cosh' method ($\circ$).
$T$ is in aribtrary unit.
\end{document}